# Revisiting the X-ray – Mass scaling relations of Early-type Galaxies with the Mass of their Globular Cluster Systems as a Proxy for the Total Galaxy Mass

Dong-Woo Kim[1], Nicholas James[1], Giuseppina Fabbiano[1], Duncan Forbes[2], and Adebusola Alabi[3]

[1] Harvard-Smithsonian Center for Astrophysics, 60 Garden Street, Cambridge, MA 02138, USA
[2] Centre for Astrophysics & Supercomputing, Swinburne University, Hawthorn VIC 3122, Australia
[3] University of California Observatories, 1156 High Street, Santa Cruz, CA 95064, USA.

May 30, 2019

## Abstract

Using globular cluster (GC) kinematics and photometry data, we calibrate the scaling relation between the total galaxy mass ($M_{TOT}$ including dark matter) and total globular cluster system mass ($M_{GCS}$) in a sample of 30 early-type galaxies (ETG), confirming a nearly linear relationship between the two physical parameters. Using samples of 83 and 57 ETGs, we investigate this scaling relation in conjunction with the previously known relations between $M_{TOT}$ and the ISM X-ray luminosity and temperature, respectively. We confirm that $M_{GCS}$ can be effectively used as a proxy of $M_{TOT}$. We further find that the $L_{X,GAS} - M_{TOT}$ relation is far tighter in the subsample of core ETGs, when compared to cusp ETGs. In core ETGs (old, passively evolving stellar systems) $M_{TOT}$ is significantly larger than the total stellar mass $M_{STAR}$ and the correlation with the hot gas properties is driven by their dark matter mass $M_{DM}$. Cusp ETGs have typically lower $L_{X,GAS}$ than core ETGs. In cusp ETGs, for a given $M_{DM}$, higher $L_{X,GAS}$ is associated with higher $M_{STAR}$, suggesting stellar feedback as an important secondary factor for heating the ISM. Using the $M_{GCS}$-$M_{TOT}$ scaling relations we compare 272 ETGs with previous estimates of the stellar-to-halo mass relation of galaxies. Our model-independent estimate of $M_{TOT}$ results in a good agreement around halo masses of $10^{12}$ $M_\odot$, but suggest higher star formation efficiency than usually assumed both at the low and at the high halo mass ends.

# 1. INTRODUCTION

The total galaxy mass ($M_{TOT}$) out to a large radius provides an observational constraint to the amount of dark matter (DM) in galaxies, a key ingredient for the formation and evolution of galaxies (e.g., see Naab & Ostriker 2017 and references therein; Somerville and Dave 2015 and references therein). Despite of its importance, accurate measurements of $M_{TOT}$ are still challenging. While dynamical masses have been measured using integral field two-dimensional spectroscopic data for a large number of early type galaxies (ETGs) (e.g., in the Atlas 3D sample; Cappellari et al. 2013), these data are limited to radii within ~1 Re (effective radius or half-light radius) where the stellar mass dominates over DM. At large radii, dynamical mass measurements are provided by the analysis of the kinematics of hundreds of globular clusters (GC) and planetary nebulae (PN) in individual galaxies (Deason et al. 2012, Alabi et al. 2017), but, so far these measurements are available only for a small number (~30) of ETGs.

A different, less direct approach for estimating $M_{TOT}$ makes use of other observational proxies. In our earlier work, we have established the X-ray luminosity of the hot ISM ($L_{X,GAS}$) and its temperature as proxies of $M_{TOT}$ (Kim & Fabbiano 2013; Forbes et al. 2017), at least for core ETGs. The near-linear relationship between the total mass of the galaxy's GC system ($M_{GCS}$) and $M_{TOT}$ (Blakeslee et al. 1997; Spitler & Forbes 2009; Hudson et al. 2014) suggests that $M_{GCS}$ could also be a proxy of $M_{TOT}$. $M_{GCS}$ would have the advantage of being available for a large sample of galaxies.

This study consists of two parts. First, we calibrate $M_{GCS}$ as a proxy of $M_{TOT}$, by comparing it with good-quality kinematics measurements of $M_{TOT}$ within 5 Re available for 30 ETGs, and with the X-ray proxies in a sample of 83 ETGs, for which X-ray and $M_{GCS}$ data exist. Second, we further study the differences between core and cusp ETGs suggested by our earlier work investigating the secondary factors responsible for heating the ISM in the low mass (low $L_{X,GAS}$) cusp ETGs.

Kim & Fabbiano (2013) and Forbes et al. (2017) noted that the $L_{X,GAS} - M_{TOT}$ relationship is particularly tight for gas-rich galaxies with core surface brightness profiles, indicating that $M_{TOT}$ is the primary factor in retaining hot ISM in these galaxies. Instead, more scatter is observed for X-ray fainter cusp galaxies, suggesting that non-gravitational effects may be at play. Besides gravitational heating during infall, the gas released into the ISM by evolved stars and supernovae is also heated by stellar and AGN feedback (Pellegrini 2011, 2012). The complex balance between these energizing processes, the depth of the galaxy's potential well, external mergers and stripping, outflows and replenishment from stellar sources determines the gas temperature, density and luminosity that is observed in the present day. However, the balance between these processes is still not fully understood.

The ETG samples we have assembled for this work are described in Section 2. In Section 3, (1) we compare $M_{GCS}$ with the total mass ($M_{TOT}$ within 5 Re) determined by GC kinematics data for 30 ETGs and (2) with the full sample of 83 ETGs, establishing that $M_{GCS}$ is a good proxy for $M_{TOT}$. In Section 4, we further investigate and discuss the X-ray scaling relationships for core and cusp ETGs. Our results are summarized in Section 5. Throughout this paper, we quote errors at the 1σ significance level.

# 2. THE ETG SAMPLE

In this section, we summarize the provenance of the three main data sets used in this study: GC photometric data and $M_{GCS}$ from Harris et al. (2013, 2017); total galaxy mass ($M_{TOT}$ within 5 Re) data, measured through GC kinematics, from Alabi et al. (2017); and X-ray data from which we derive $L_{X,GAS}$ and $T_{GAS}$ (the gas temperature) from various sources described below. The data sets and their properties relevant to this study are summarized in Table 1, which contains basic information about the galaxies, the X-ray data, $M_{GCS}$, and $M_{TOT}$.

## 2.1. Globular Cluster Photometric Data

The photometric GC data used in this study come from the catalog of Harris et al. (2013), who compile the number of GCs in 341 ETGs from the literature. Of these ETGs, 83 have X-ray data (see Section 2.3). For completeness, we list the sources of GC data (taken from Harris et al. 2013) for individual galaxies in Table 1. To correct for incompleteness, counts are generally extrapolated out to the full extent of the galaxy based on their observed radial profile and extrapolated to lower magnitudes by fitting a GC luminosity function (typically Gaussian in shape) to the number and luminosity of observed GCs. Harris et al. additionally determine $M_{GCS}$ for each galaxy in their catalog. Harris et al. determine the total GC V-band luminosity in each galaxy using the galaxies' V-band magnitudes and the GC luminosity function described in Jordan et al. (2006) and Vesperini (2010). To derive $M_{GCS}$, they then scale total GC luminosity to total GC mass using a mass-to-light ratio of 2.

Given that the inhomogeneous nature of the GC data, there may be unknown systematic biases or selection effects. Therefore, in section 3, we test the validity of $M_{GCS}$ as a proxy or $M_{TOT}$ by comparing with (1) the kinematically determined total mass ($M_{TOT}$ within 5Re), (2) the hot gas X-ray luminosity ($L_{X,GAS}$) and (3) the hot gas temperature ($T_{GAS}$).

## 2.2. Globular Cluster Spectroscopic Data: Kinematic Mass Measurements ($M_{TOT}(5Re)$)

Alabi et al. (2017) use spectroscopically determined line-of-sight velocities of GCs to determine the total mass (baryonic + dark matter) within five effective radii, $M_{TOT}(5Re)$, for 32 ETGs of which 30 (with the exclusion of NGC 2974 and NGC 4474) have X-ray measurements. Alabi et al. assume a GC power-law density distribution within a given galaxy, and a power-law profile for the galaxy's gravitational potential. They calculate the power-law slope of the GC density distribution for each galaxy based on its stellar mass, using an empirical relationship determined from GC density profiles in the literature. They also calculate the power-law slope of each galaxy's gravitational potential based on its stellar mass, using the relationship determined from the cosmological simulations of Wu et al. (2014). The mass of each galaxy is assumed to have a pressure-supported component and a significantly smaller rotationally supported component, which are calculated separately based on the GC radial velocities and the assumed radial profiles of GCs and gravitational potential. $M_{TOT}$ is the combination of the pressure-supported mass and the rotationally supported mass. The estimated uncertainty on $M_{TOT}(5Re)$ varies with the total number of tracers used, such that when the number of GCs is $N_{GC} > 100$, typical uncertainty is ~0.1 dex. For galaxies with $N_{GC}$ ~ 70 and < 40, typical uncertainties on $M_{TOT}(5Re)$ are ~0.2 and ~0.25, respectively.

## 2.3. X-ray Data

We have assembled a sample of 83 ETGs with measurements of the X-ray luminosity from the hot gas ($L_{X,GAS}$) from several sources. The bulk of the X-ray data used in this work are Chandra data from Kim & Fabbiano (2015; hereafter KF15) and Boroson, Kim, & Fabbiano (2011; hereafter BKF11). Both papers present $T_{GAS}$ and $L_{X,GAS}$ (within the 0.3-8.0 keV energy range) of each galaxy in their respective samples, after removing the contributions to each galaxy's X-ray luminosity of low mass X-ray binaries (LMXBs), active binaries (ABs) and cataclysmic variables (CVs), and the active galactic nucleus (AGN), if present. The BKF11 sample consists of 30 nearby non-cD (core dominant) early-type galaxies, while the KF15 sample consists of the 60 early-type galaxies in the volume-limited ATLAS$^{3D}$ sample (Cappellari et al. 2011) that had Chandra ACIS observations longer than 10 ksec. There is considerable overlap between the two data sets, with the combined sample consisting of 48 galaxies.

We supplement the KF15 and BKF11 data with additional Chandra data from Goulding et al. (2016, hereafter G16), adding 9 galaxies in total. The G16 sample consists of the 33 galaxies within the MASSIVE data set (a survey of the 116 most massive early-type galaxies within 108 Mpc) that have archival Chandra observations. Following the procedure of KF15 and BKF11, G16 remove the contribution of X-ray emission by point sources, such that their reported luminosities are contributed to solely by gas emission. G16 present their X-ray luminosities within the 0.3-5.0 keV energy range. We converted these luminosities to the 0.3-8.0 keV energy range used by BKF11 and KF15 using PIMMS[1] (Portable, Interactive Multi-Mission Simulator).

We additionally supplement this X-ray data set with ROSAT data from O'Sullivan et al. (2003, hereafter OPC03) and O'Sullivan et al. (2001, hereafter OFP01). In total we use data for 4 galaxies from OPC03 and 20 from OFP01. We applied corrections to the OPC03 and OFP01 luminosities to convert them to the same 0.3-8.0 keV energy band as the rest of the X-ray data that we use, and to remove the contribution of point source emission, following BKF11. OFP01 do not include any estimate for the uncertainty in their luminosities, so we assume fractional uncertainties of 50% for the OFP01 luminosities, consistent with the typical value seen in the similarly derived OPC03 data.

Finally, we add Sombrero (NGC 4594, S0) and CenA (NGC 5128, S0 pec) to our sample because they have GC spectroscopic data (see section 2.3). Their hot gas properties were measured with Chandra data by Li & Wang (2013) and Kraft et al. (2003), respectively. In total, our X-ray sample consists of 83 ETGs.

Table 1 also provides a classification of our sample galaxies as 'core' or 'cusp'. Considering the so-called E-E dichotomy (e.g., Kormendy & Ho 2013), we further split the sample up by nuclear profile, into "core" ETGs, which have nuclear surface brightness profiles that flatten out towards the center, and power-law or "cusp" ETGs, the surface brightness profiles of which continue to increase up to the resolution limit. Core ETGs tend to be luminous, slow rotators, while cusps are less luminous and more rapidly rotating. Additionally, core ETGs tend to have boxy isophotes, while those of cusp ETGs tend to be more disky. In general, core ETGs consist of a homogeneous sample of pure elliptical galaxies with no recent star formation, while cusp ETGs can be heterogeneous with respect to recent star formation, galaxy shape, and rotation. Throughout this work, assignment of "core" or "cusp" is based on the results of Lauer et al. (2005, 2007), Cote et al. (2006), Hopkins et al (2009a, b), Richings et al. (2011) and Krajnović et al. (2013).

---

[1] http://cxc.harvard.edu/toolkit/pimms.jsp

### 2.3.1 Validation of the aggregated X-ray sample

We investigated our aggregated X-ray data set by revisiting the well-studied $L_{X,GAS} - L_K$ and $L_{X,GAS} - T_{GAS}$ scaling relationships of the hot ISM (e.g., BKF11, KF15, G16). Figure 1 plots the ISM X-ray luminosity of each galaxy in our aggregated sample against the K-band luminosity which is a good proxy for stellar mass for ETGs (e.g., Bell et al. 2003). As in KF15, we have applied a bisector linear regression method (Akritas & Bershady 1996) and estimated the corresponding error by bootstrap resampling[2]. We have tested our results with the new Bayesian approach with MCMC simulations given in Kelly (2007) and found our results are consistent within a 1 sigma uncertainty for all relations we studied in the paper. We also note that the range of a 1 sigma percentile given by the MCMC run is always within the RMS scatter (the shaded area in all figures). We also used the Pearson and Spearman correlation tests from the scipy statistics package (http://www.scipy.org) to estimate the p-value for the null hypothesis. The results are summarized in Tables 2 and 3 for $L_{X,GAS}$–$L_K$ and $L_{X,GAS}$–$T_{GAS}$, respectively. In this sample of 83 galaxies with $L_{X,GAS}$, we find a best fit power-law slope of $2.69 \pm 0.23$. This is slightly flatter than the slope found by KF15 ($2.98 \pm 0.36$ for a full sample), but consistent within the $1\sigma$ error.

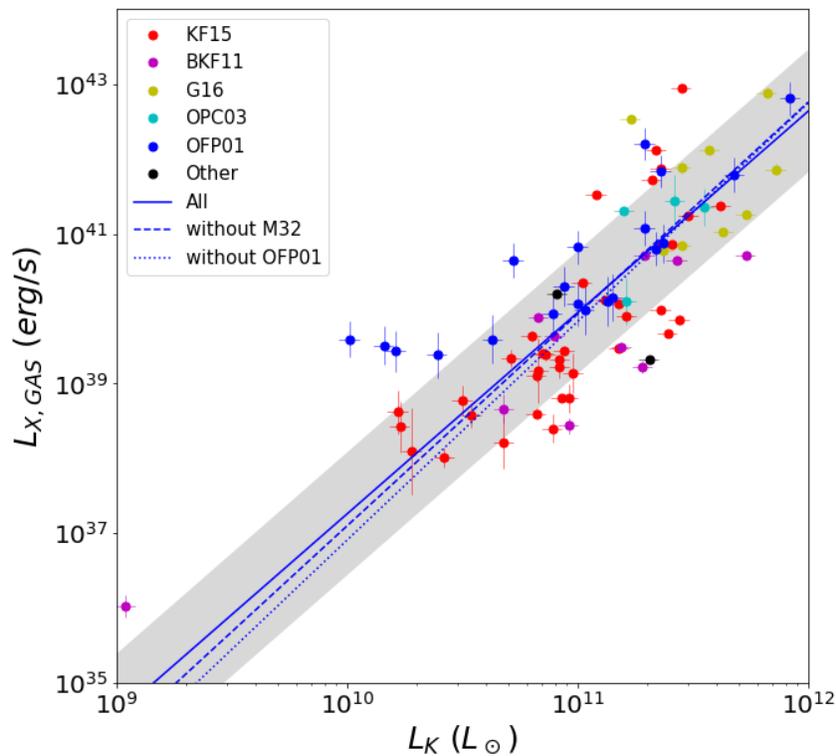

Fig. 1. ISM X-ray luminosity is plotted against K-band luminosity (a proxy for stellar mass) for the 83 galaxies in our aggregated X-ray data set. Data points are color-coded based on the source of the X-ray data. The solid line is the best fit relation for the entire sample and the grey shade indicates the RMS scatter from the best fit. The dashed (dotted) line is without M32 (OFP01).

---

[2] Python version from http://home.strw.leidenuniv.nl/~sifon/pycorner/bces

Excluding M32 (NGC 221, the point at the lower left corner), we best-fit parameters entirely consistent with KF15.

While data points from difference sources are in general consistent with each other, $L_{X,GAS}$ from OFP01 seems to be slightly higher than those from Chandra measurements for a given $L_K$, particularly at the low $L_K$. However, the three galaxies (NGC 4387, NGC 4458, NGC 4550) with the lowest $L_K$ among OFP01 may have a measurement error (e.g., the contributions from LMXBs and AGN were unresolved and incorrectly subtracted). Also, we cannot exclude a selection effect (faint galaxies might have been excluded in the ROSAT sample). We therefore repeated the linear regression without OFP01 and obtain similar parameters, except that the intercept is slightly lower, but still consistent within the error (see Table 2). We note that only one (NGC 4387) of the three galaxies with the lowest $L_K$ has a cusp profile (the other two being intermediate/unknown) so that they do not affect our analyses presented for the core and cusp subsamples throughout this paper.

```
Table 2. Best fit parameters for the Lx,GAS - LK relation
================================================================================
sample        # of galaxies   slope error   intercept error    RMS      p-value
                                                              (dex)  pearson  spearman
--------------------------------------------------------------------------------
All               83          2.69  0.23    10.36  2.52       0.85   3.3e-18  1.0e-18
Without M32       82          2.83  0.24     8.80  2.70       0.86   1.2e-15  6.0e-18
Without OFP01     63          2.92  0.32     7.71  3.52       0.81   2.2e-16  1.2e-16
================================================================================
```

Figure 2 plots the ISM luminosity vs. temperature for each galaxy in the sample for which there is an available temperature measurement (and thus does not include the OFP01 data, as well as some galaxies from the other data sets with unconstrained temperatures). The aggregated data set consists of 57 galaxies. We find a best-fit power-law slope of $5.06 \pm 0.32$ (see Table 3), in agreement with the $5.39 \pm 0.60$ slope found by KF15 (for a full sample).

Given the agreement between the scaling laws we obtain from our larger sample with previous results from more homogeneous selections, we conclude that the present sample of ETGs is representative of the X-ray properties of ETGs.

```
Table 3. Best fit parameters for the Lx,GAS - TGAS relation
================================================================================
sample        # of galaxies   slope error   intercept error    RMS      p-value
                                                              (dex)  pearson  spearman
--------------------------------------------------------------------------------
All               57          5.06  0.32    41.70  0.12       0.67   7.9e-17  2.3e-14
================================================================================
```

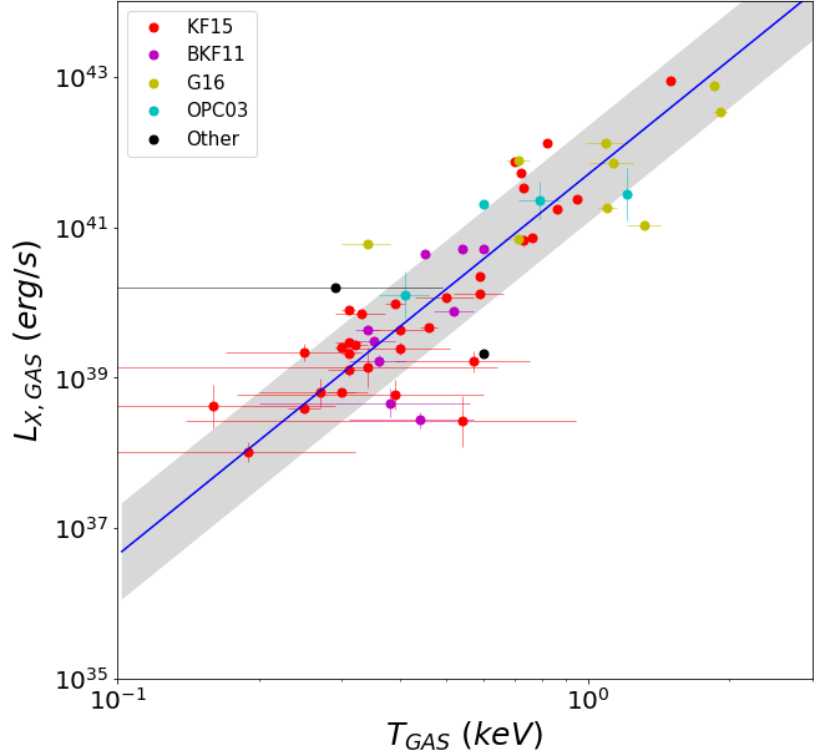

Fig. 2. ISM X-ray luminosity is plotted against ISM temperature for the 57 galaxies with measured temperatures in our aggregated X-ray data set. Data points are color-coded based on the source of the X-ray data.

## 3. $M_{GCS}$ AS PROXY OF $M_{TOT}$

As discussed in Section 1, previous studies have shown a linear relationship between the (empirically determined) total mass ($M_{TOT}$) of a galaxy and the total mass ($M_{GCS}$) of its GC system (Harris et al. 2013, 2015, 2017). Here we test and calibrate the $M_{GCS} - M_{TOT}$ relation by means of a direct comparison with kinematically derived $M_{TOT}$ (Section 3.1). We then extend this comparison to include the X-ray luminosity from the hot gaseous halos in these galaxies $L_{X,GAS}$ and the temperature of the hot gas from X-ray spectral fits $T_{GAS}$ (Section 3.2).

### 3.1. Comparison of $M_{GCS}$ with kinematically measured $M_{TOT}$

Fig. 3 shows the relation between $M_{TOT}$ and $M_{GCS}$, where $M_{TOT}$ was measured from GC kinematics data by Alabi et al. (2017). Alabi et al. measured the total mass, $M_{TOT}(5R_e)$, within five effective radii ($R_e$) where most GCs are found. Figure 3 (left) displays this relationship for the sample of 30 galaxies with both Alabi et al. (2017) $M_{TOT}(5R_e)$ data and Harris et al. (2013) $M_{GCS}$ data. Alabi et al. (2017) also provided $M_{200} = M_{TOT}(R_{200})$ by extrapolating their measurement to $R_{200}$. In Figure 3 (right), we show the $M_{200} - M_{GCS}$ relation. The results of the analysis of the correlations shown in Figure 3 are summarized in Table 4.

We find a strong, close-to-linear relationship between $M_{TOT}(5Re)$ and $M_{GCS}$, with a power-law slope of $0.85 \pm 0.06$, and a root-mean square (RMS) deviation of 0.27 dex. The corresponding p-values for the null hypothesis by the Pearson and Spearman tests are $10^{-9}$ and $2 \times 10^{-8}$, respectively. The relationship between $M_{TOT}(5Re)$ and $M_{GCS}$ for the core subsample (see Section 2.3) has a slope of $0.87 \pm 0.09$ and a similar RMS deviation to that of the full sample.

$$\log (M_{TOT}(5Re) / 10^{11.7} M_\odot) = 0.85 \pm 0.06 \times \log (M_{GCS} / 10^{8.5} M_\odot) \text{ for the full sample}$$
$$\log (M_{TOT}(5Re) / 10^{11.7} M_\odot) = 0.87 \pm 0.09 \times \log (M_{GCS} / 10^{8.5} M_\odot) \text{ for the core subsample}$$

The slope for the cusp subsample is similar but not as well constrained, with a power-law slope of $0.83 \pm 0.41$, and p-value of 0.19 and 0.06 for the Pearson and Spearman tests, respectively. The p-value is considerably higher than those of the full and core samples, indicating a weak or no correlation. This could be partly because the sample is small and because the dynamic range is narrow, e.g., $M_{GCS}$ spanning one decade in the cusp subsample, compared to two decades in the core subsample.

For the full sample and the core subsample, the relationships between $M_{200}$ and $M_{GCS}$ are similar to the above relations, except that the slope is entirely consistent with being linear.

$$\log (M_{TOT}(R_{200}) / 10^{13.1} M_\odot) = 0.99 \pm 0.07 \times \log (M_{GCS} / 10^{8.5} M_\odot) \text{ for the full sample}$$
$$\log (M_{TOT}(R_{200}) / 10^{13.1} M_\odot) = 1.01 \pm 0.1 \times \log (M_{GCS} / 10^{8.5} M_\odot) \text{ for the core subsample}$$

Again, the cusp subsample has a large error in slope ($0.84 \pm 0.4$) and the corresponding p-values are 0.18 and 0.04 for Pearson and Spearman tests, respectively.

Since the correlation parameters are statistically identical in the full sample and the core/cusp subsamples (Table 4), we use the relation of the full sample in this paper. We consider the error in $M_{GCS}$ and the uncertainty in the $M_{GCS}$-$M_{TOT}(5Re)$ relation (the RMS scatter) to calculate the error of $M_{TOT}$ when scaled from $M_{GCS}$.

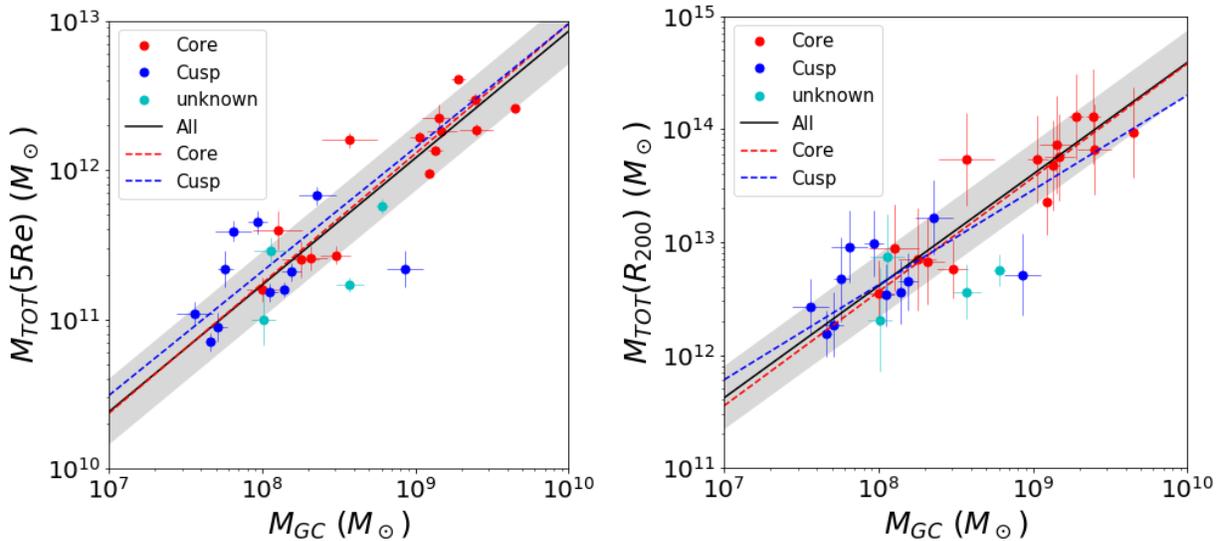

Fig. 3. $M_{TOT}$ is plotted against $M_{GCS}$ for the Alabi et al. (2017) sample. $M_{TOT}$ is determined by GC kinematics data within (left) five effective radii (5Re) and (right) $R_{200}$. The red and blue points indicate the core and cusp subsamples, respectively. The three lines are the best fit regression lines for all (solid black), core (dashed red) and cusp (dashed blue) ETGs.

```
Table 4. Best fit parameters for the MTOT - MGCS relation
==========================================================================================
sample     # of galaxies  slope  error  intercept  error   RMS      p-value
                                                           (dex)  pearson  spearman
------------------------------------------------------------------------------------------
with MTOT(5Re)
All            30         0.85   0.06   4.43       0.52    0.27   1.1e-09  1.9e-08
core           15         0.87   0.09   4.28       0.83    0.20   5.3e-06  4.9e-06
cusp           11         0.83   0.41   4.68       3.23    0.32   0.19     0.06

with MTOT(R200)
All            30         0.99   0.07   4.69       0.58    0.32   2.8e-09  1.4e-07
core           15         1.01   0.10   4.48       0.91    0.23   5.1e-06  1.0e-05
cusp           11         0.84   0.40   5.90       3.18    0.33   0.18     0.04
==========================================================================================
```

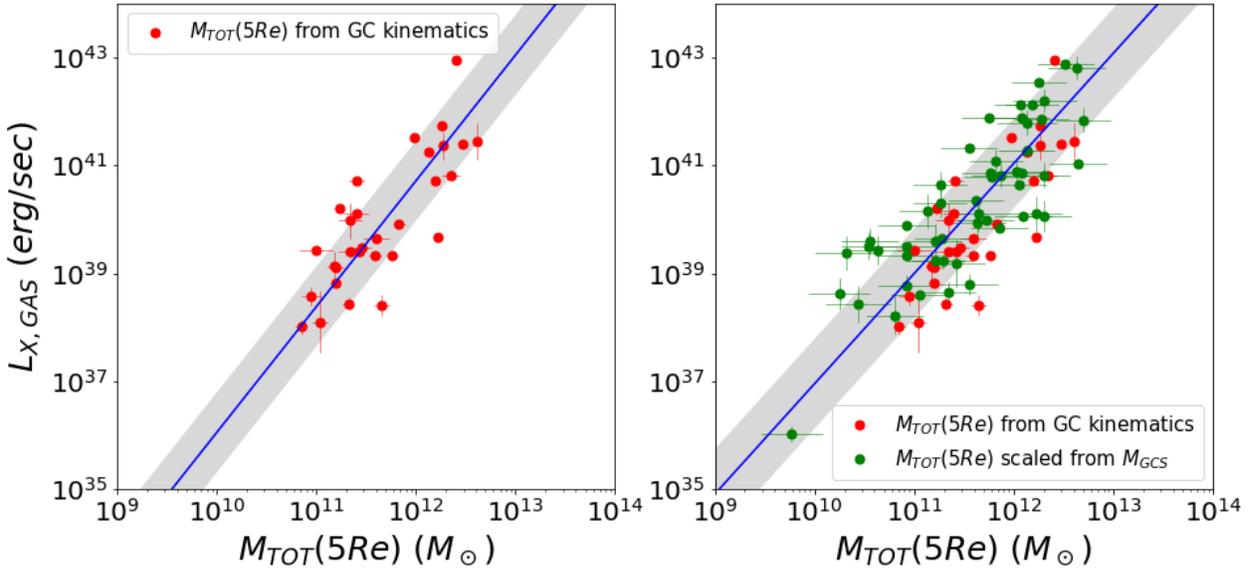

Fig. 4. The $L_{X,GAS} - M_{TOT}$ relationship. In the left plot, the 30 ETGs with kinematically determined mass measurements (within five effective radii) are used. In the right, the same data points are plotted, along with the 53 additional galaxies with total masses estimated based on $M_{GCS}$.

### 3.2. Comparisons of $M_{TOT}$ from GC kinematics and $M_{TOT}$ from $M_{GCS}$ with $L_{X,GAS}$ and $T_{GAS}$

The X-ray luminosity of the hot ISM ($L_{X,GAS}$) is known to correlate with the total mass (Kim & Fabbiano 2013; Forbes et al. 2017). Using the total galaxy masses both directly measured from the GC kinematics and scaled from $M_{GCS}$ (as described in Sections 3.1), we are able to study the $L_{X,GAS}$

− $M_{TOT}$ scaling relationship in a larger data set (83 ETGs) than ever before. Also, comparing the $L_{X,GAS} - M_{TOT}$ relations determined with two sets of $M_{TOT}$, we can test the validity of the relation described in section 3.1. To estimate $M_{TOT}$ by scaling from $M_{GCS}$, we apply the $M_{TOT}$-$M_{GCS}$ relation for the full sample (section 3.1).

Fig. 4 shows the $L_{X,GAS} - M_{TOT}$ relations we obtain using GC kinematics (left panel; 30 ETGs), and both kinematics and the $M_{TOT} - M_{GCS}$ relation (right panel; 83 ETGs). The results of the analysis of these relations are summarized in Table 5. For the sample of 30 galaxies with kinematic $M_{TOT}$, we find a best-fit power-law slope of $2.33 \pm 0.24$, with an RMS deviation of 0.74. This slope is lower than the slope of $3.13 \pm 0.32$ found by Forbes et al. (2017) using the same data set. This discrepancy is primarily attributed to the new effective radii used by Alabi et al. (2017) for their mass measurements, which differ from those used by Forbes et al. (2017). We could reproduce the previous relationship with the old Re values in Forbes et al. (2017). Using ultra deep optical observations, recent studies (e.g., Ferrarese et al. 2006; Spavone et al. 2017) indeed suggest the effective radii of ETGs are larger than the previous values which were commonly used (e.g., RC3, 2MASS).

For the 83 ETGs with both X-ray and total mass data (30 kinematically determined, and 53 scaled from $M_{GCS}$ using the relations of Table 4), we find a best fit power-law slope of $2.04 \pm 0.14$ for the $L_{X,GAS} - M_{TOT}(5Re)$ scaling relation, with an RMS deviation of 0.8. This relation is consistent (1.04$\sigma$ difference) with the relation measured using the 30 ETGs with kinematically determined $M_{TOT}(5Re)$. We note that M32 (at the lower left corner in the right panel of Fig 4.) follows the overall relation. As in section 2.1 (Fig 1), excluding M32 does not change the relationship, within statistics.

The difference between core and cusp ETGs becomes more dramatic (in comparison with Section 3.1, Fig. 3), when $L_{X,GAS}$ (i.e. the properties of the hot gas) is considered. The 35 ETGs with a confirmed core profile (left panel of Figure 5 and Table 5), exhibit a tight correlation and have a slightly lower RMS deviation of 0.7. We note that the two datasets with kinematically determined $M_{TOT}$ (red) and those scaled from $M_{GCS}$ (green) lie in the same parameter space in the $T_{GAS} - M_{TOT}$ space.

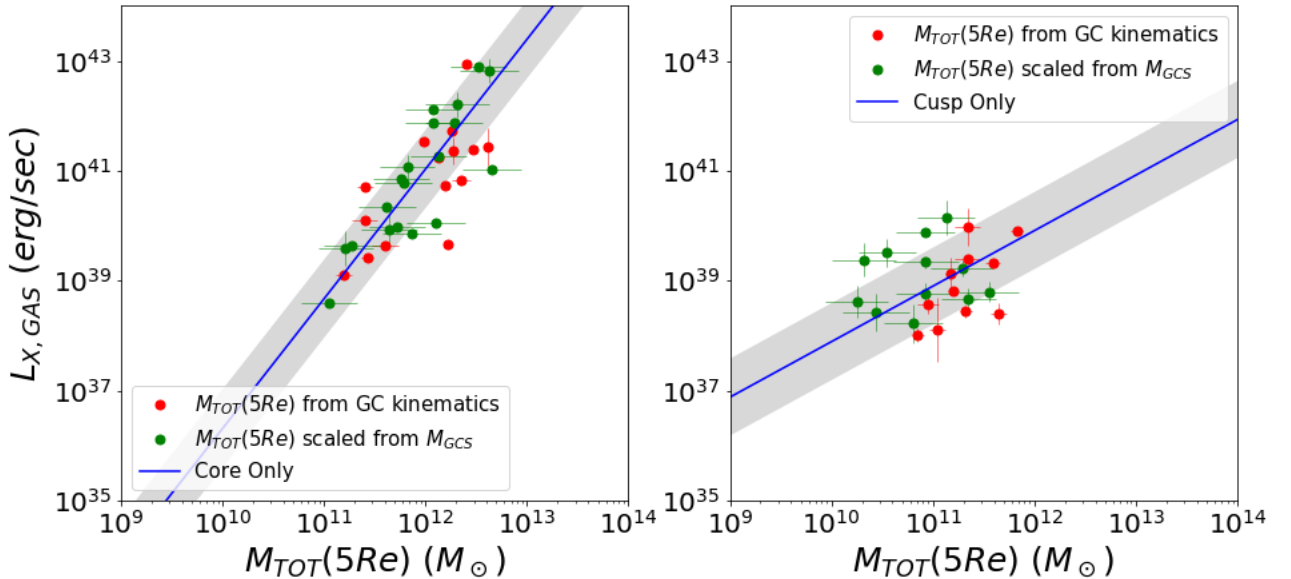

Fig. 5. The $L_{X,GAS}$ − $M_{TOT}$ relationship. Left: the 35 galaxies with a core profile. Right: the 21 galaxies with a cusp profile.

The lower mass, cusp ETGs show no correlation (right panel of Figure 5 and Table 5). The slope of the $L_{X,GAS}$ – $M_{TOT}$(5Re) relationship for these galaxies is unconstrained (1.01 ± 1.12), with a p-value of 0.9. The small dynamic range of the cusp subsample might have caused the observed scatter. The two data with kinematically determined $M_{TOT}$ (red) and those scaled from $M_{GCS}$ (green) may be offset as the green points have lower $M_{TOT}$ than the red points while they occupy similar $L_{X,GAS}$ ranges. This may indicate that the uncertainty in $M_{TOT}$ when scaled from $M_{GCS}$, or that those kinematic $M_{TOT}$ sample preferentially selected a bigger system than the photometric sample. Based on the 2-dimensional Kolmogorov–Smirnov test (Fasano & Franceschini 1987), the probability that two sub-samples are originated from the same parent population is 0.05, which makes both possibilities open as the commonly used p-value limit is 0.05. We further discuss in section 4 the implications of these results.

```
Table 5. Best fit parameters for the Lx,GAS - MTOT(5Re) relation
================================================================================
sample         # of galaxies   slope error   intercept error    RMS       p-value
                                                               (dex)  pearson  spearman
--------------------------------------------------------------------------------
Mass(kinematics)   30          2.33  0.24    12.75  2.83       0.74   8.1e-08  4.9e-07
All                83          2.04  0.14    16.55  1.66       0.83   6.8e-19  1.6e-18
core               35          2.36  0.22    12.71  2.57       0.71   2.8e-08  6.7e-08
cusp               21          1.01  1.12    27.79 12.39       0.70   0.96     0.85
================================================================================
```

In summary, the $L_{X,GAS}$ − $M_{TOT}$ correlation is tight in the sample with kinematically determined $M_{TOT}$ as well as in the full sample including those scaled from $M_{GCS}$ and the two relations are consistent within 1σ. This trend is more obvious among the core subsample where $L_{X,GAS}$ is directly related to the total mass.

Using the $M_{TOT}$-$M_{GCS}$ relation, we next expand the comparison to the $T_{GAS}$ − $M_{TOT}$ scaling relationship to a larger data set (57 ETGs) than ever before. We note that while both $T_{GAS}$ and $L_{X,GAS}$ can be determined by the same X-ray observation, $T_{GAS}$ is determined by the spectral shape and $L_{X,GAS}$ by the normalization. Therefore, $T_{GAS}$ and $L_{X,GAS}$ are independent quantities. Because the gas temperature is expected to be directly linked to the gravitational potential or the virial mass (e.g., Navarro et al. 1995; Sanderson et al. 2003), comparing the $T_{GAS}$ − $M_{TOT}$ relations determined with two sets of $M_{TOT}$, we can further test the validity of the $M_{TOT}$ − $M_{GCS}$ relation.

Fig. 6 shows the $T_{GAS}$ − $M_{TOT}$ relations we obtain using GC kinematics (left panel; 26 ETGs), and both kinematics and the $M_{TOT}$− $M_{GCS}$ relation (right panel; 57 ETGs). The results of the analysis of these relations are summarized in Table 6. For the sample of 26 galaxies with kinematic $M_{TOT}$, we find a best-fit power-law slope of 0.44 ± 0.04, with an RMS deviation of 0.12. For the 57 ETGs with both X-ray and total mass data (26 kinematically determined, and 31 scaled from $M_{GCS}$ using the relations of Table 4), we find a best fit power-law slope of 0.44 ± 0.03 for the $T_{GAS}$ − $M_{TOT}$(5Re) scaling relation, with an RMS deviation of 0.16. This relation is statistically identical with the relation measured using the 26 ETGs with kinematically determined $M_{TOT}$(5Re).

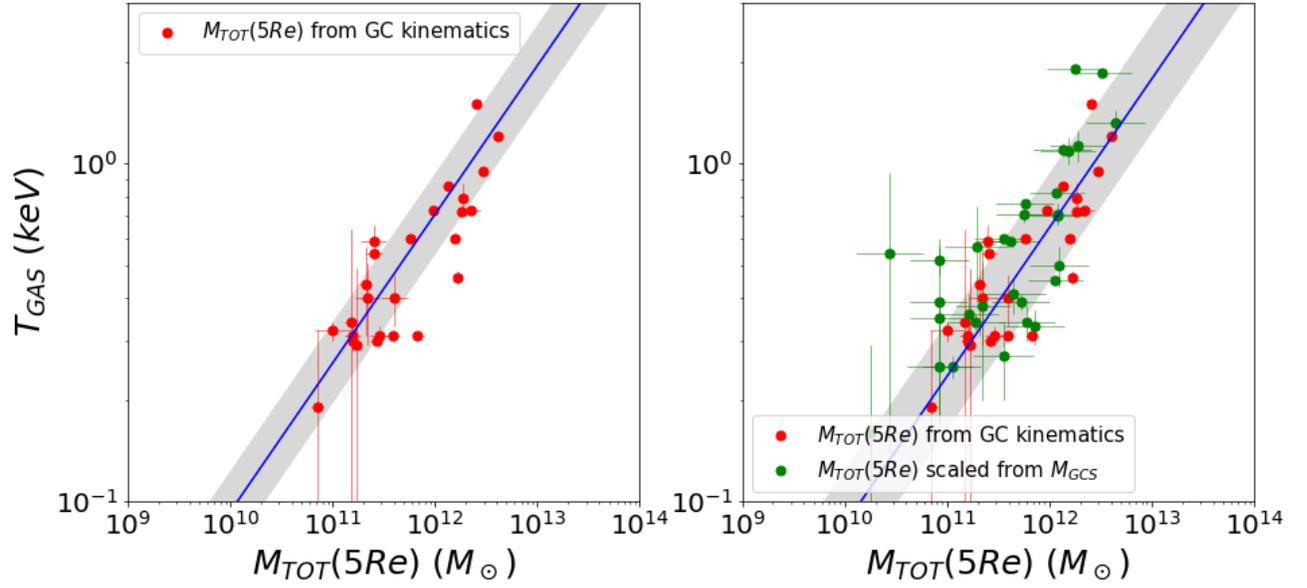

Fig. 6. The $T_{GAS} - M_{TOT}$ relationship. In the left plot, the 26 ETGs with kinematically determined mass measurements (within five effective radii) are used. In the right, the same data points are plotted, along with the 31 additional galaxies with total masses estimated based on $M_{GCS}$.

Separating the core and cusp subsamples, we find that the 29 ETGs with a confirmed core profile (left panel of Figure 7) exhibit a tight correlation, but the 15 cusp ETGs show no correlation (right panel of Figure 7). The slope of the core sample is well constrained (0.52 ± 0.05), while the slope of the cusp sample is unconstrained (0.66 ± 0.58) with a p-value of 0.5-0.8. Again, it is somewhat uncertain that the lack of the correlation in the cusp subsample is real, because of the small dynamic range.

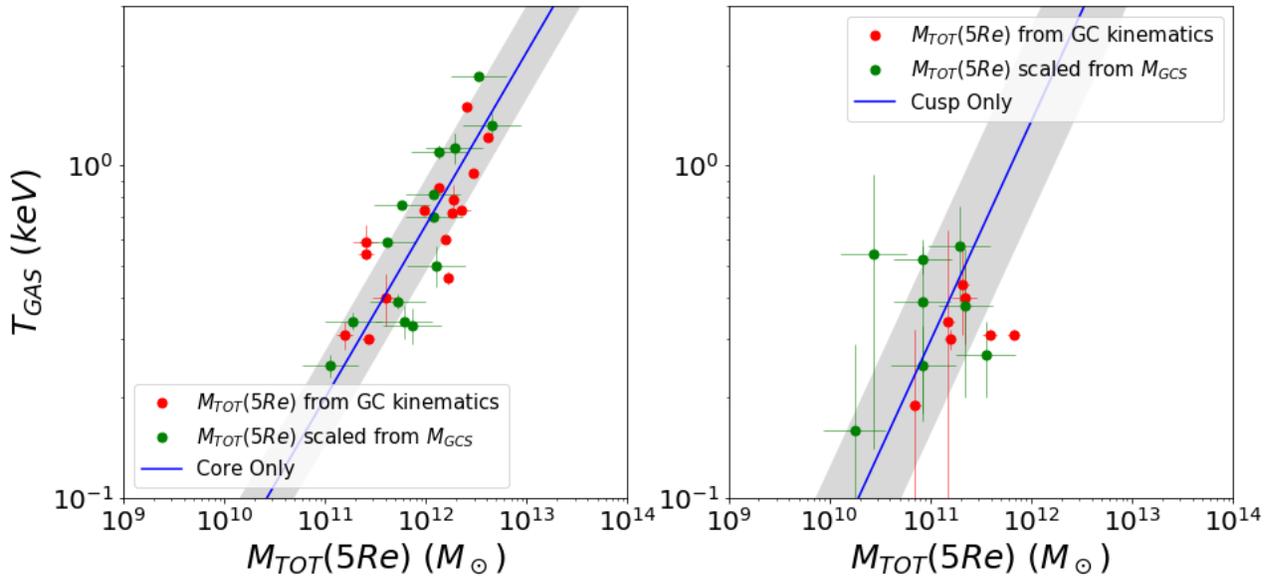

Fig. 7. The $T_{GAS} - M_{TOT}$ relationship. Left: the 29 galaxies with a core profile. Right: the 15 galaxies with a cusp profile.

In the core subsample, the two datasets with kinematically determined $M_{TOT}$ (red) and those scaled from $M_{GCS}$ (green) share the same parameter space in the $T_{GAS} - M_{TOT}$ space. In the cusp subsample, the green points may have lower $M_{TOT}$ than the red points while they occupy similar $T_{GAS}$ ranges. Based on the 2-dimensional Kolmogorov–Smirnov test (Fasano & Franceschini 1987), the probability that two sub-samples are originated from the same parent population is 0.2, indicating that we cannot statistically separate them.

In summary, the $T_{GAS} - M_{TOT}$ correlation is tight in the sample with kinematically determined $M_{TOT}$ as well as in the full sample including those scaled from $M_{GCS}$ and the two relations are identical. This trend is more obvious in the core subsample than in the cusp subsample.

```
Table 6. TGAS - MTOT(5Re) relation
=============================================================================
sample        # of galaxies   slope error   intercept error    rms       p-value
                                                              (dex)  pearson   spearman
-----------------------------------------------------------------------------
Mass(kinematics)  26          0.44  0.04    -5.43  0.43       0.12   2.9e-08   6.5e-07
All               57          0.44  0.03    -5.47  0.41       0.16   2.5e-12   2.8e-12
core              29          0.52  0.05    -6.42  0.62       0.14   8.1e-08   9.0e-08
cusp              15          0.66  0.58    -7.79  6.52       0.28   0.49      0.82
=============================================================================
```

## 4. DISCUSSIONS

The analysis of our representative sample of ETGs strongly supports the use of $M_{GCS}$ as a proxy for the $M_{TOT}$ of ETGs. In particular, using the 30 galaxies in the sample with both $M_{GCS}$ and $M_{TOT}$ measured independently (Section 3.1), we find a strong correlation between $M_{TOT}$ (r < $R_{200}$ or r < 5Re) and $M_{GCS}$. Further comparing $L_{X,GAS}$ (and $T_{GAS}$) with $M_{TOT}$ (either kinematically determined or scaled from $M_{GCS}$) in a sample of 83 ETGs, we also find strong correlations (Section 3.2). These correlations persist for the subsample of core ETGs, making $L_{X,GAS}$ (and $T_{GAS}$) a good proxy of $M_{TOT}$ in these galaxies. This is in agreement with the conclusions (based on significantly smaller ETG samples) of Kim & Fabbiano (2013, 2015) and Forbes et al. (2017), that gravity is the dominant factor for the retention of hot gas in core ETGs. The lack of correlation in the subsample of cusp ETGs suggests that factors other than total mass are dominant in determining the retention of hot ISM in these galaxies, in agreement with the conclusion of Kim & Fabbiano (2015). Using our sample of 83 ETGs (both core and cusp), we re-discuss the $L_{X,GAS} - M_{TOT}$ relation and its dependence of gravity and feedback (Section 4.1). Using these scaling relations, and the sample of 272 ETGs with $M_{GCS}$ or kinematical $M_{TOT}$ available in the literature, we revisit the total galaxy mass – stellar mass fraction relation of ETGs (Section 4.2).

### 4.1 Exploring the Differences Between Core and Cusp Es

The analysis of our representative sample of ETGs supports the use of $M_{GCS}$ as a proxy for the $M_{TOT}$ of ETGs. In particular, using the 30 galaxies in the sample with both $M_{GCS}$ and $M_{TOT}$ measured independently (Section 3.1), we find a strong correlation between $M_{TOT}$ (r < 5Re) and $M_{GCS}$. Comparing $L_{X,GAS}$ ($T_{GAS}$) with either $M_{TOT}$ or the combination of $M_{TOT}$ and $M_{GCS}$ in a sample of 83 (57) ETGs, we also find strong correlations (Section 3.2). These correlations persist for the subsample of core ETGs, making $L_{X,GAS}$ and $T_{GAS}$ a good proxy of $M_{TOT}$ in these galaxies. This result is in agreement with the conclusions (based on significantly smaller ETG samples) of Kim & Fabbiano (2013, 2015) and Forbes et al. (2017) that gravity is the dominant factor for the retention of hot gas in core ETGs. In the subsample of cusp ETGs the correlations are weak or lacking, suggesting that factors other than total mass may determine the retention of hot ISM (Kim & Fabbiano 2015).

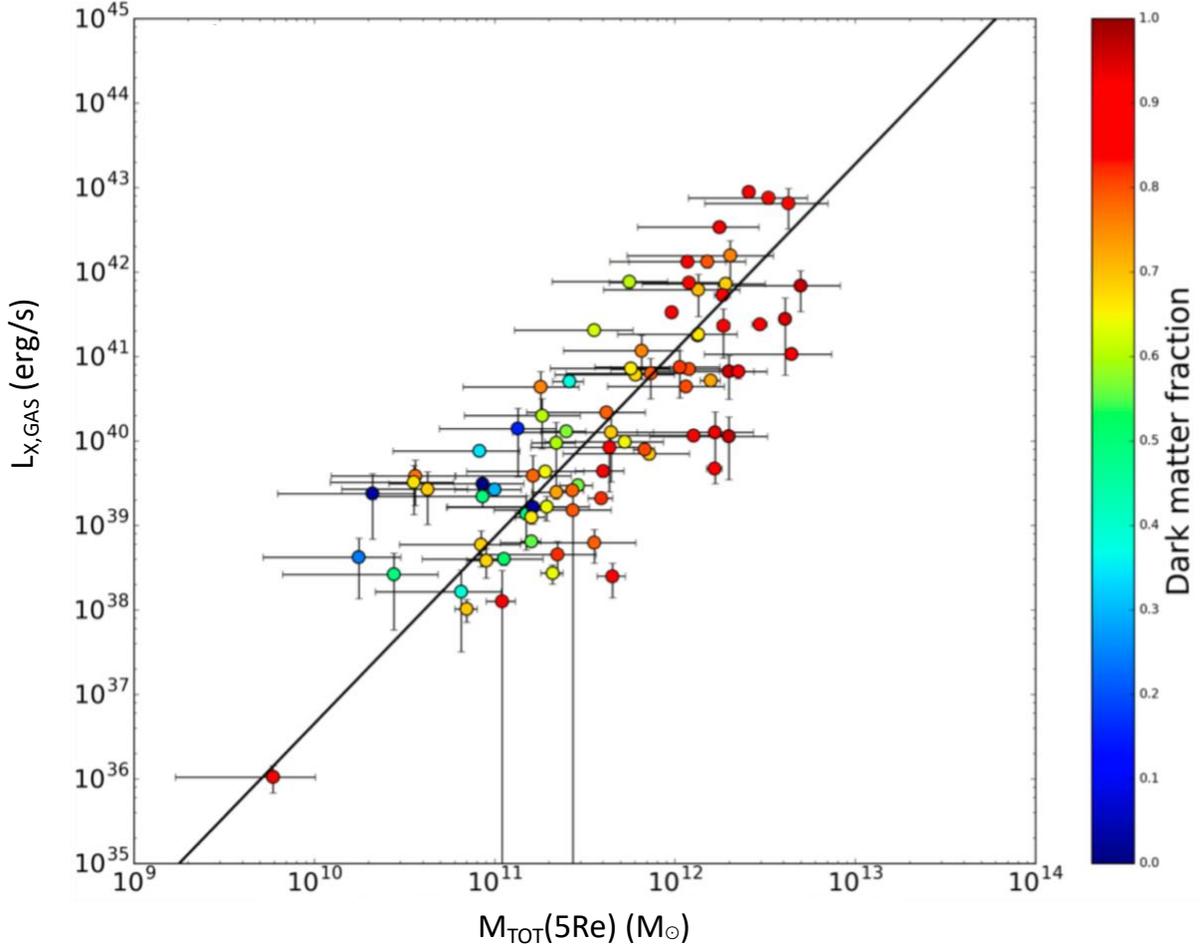

Fig. 8. The $L_{X,GAS} - M_{TOT}$ relationship, for full sample. This is the same as the top right panel of Figure 4, but color-coded according to dark matter fraction ($f_{DM}$).

Using our sample of 83 ETGs (both core and cusp), we re-discuss here the $L_{X,GAS} - M_{TOT}$ relation and its dependence on gravity and feedback. We investigated average stellar age, flattening and rotation as a third variable in the correlation (gravity or $M_{TOT}$ being the primary factor), but we could not find any systematic effect from these quantities. Instead, we found a noticeable trend with dark matter fraction ($f_{DM}$) or stellar mass fraction ($f_{STAR}$), which suggests that stellar feedback plays a critical role in the cusp subsample. This is illustrated in Figure 8, where we color-code each point according to the dark matter fraction ($f_{DM}$) we estimate for that galaxy. We estimate $M_{STAR}$ from the K-band luminosity assuming $M_{STAR}/L_K = 1\ M_\odot/L_\odot$. The uncertainty in the mass-light ratio can result in an error of ~0.1 dex in $M_{STAR}$ (e.g., Bell et al. 2003). Since in our ETG sample the gas mass is one or two orders of magnitude smaller than $M_{STAR}$, we can write $M_{TOT}$ = $M_{DM}$ + $M_{STAR}$ and $f_{DM}$ + $f_{STAR}$ = 1, from which we estimated $f_{DM}$ = $M_{DM}/M_{TOT}$. Given the uncertainties in $M_{GCS}$ and its conversion to $M_{TOT}$, we find four galaxies with negative $M_{DM}$, which we excluded from the following analyses. If, alternatively, we set $M_{DM}$=0 ($f_{STAR}$=1) for these four galaxies, our results do no change.

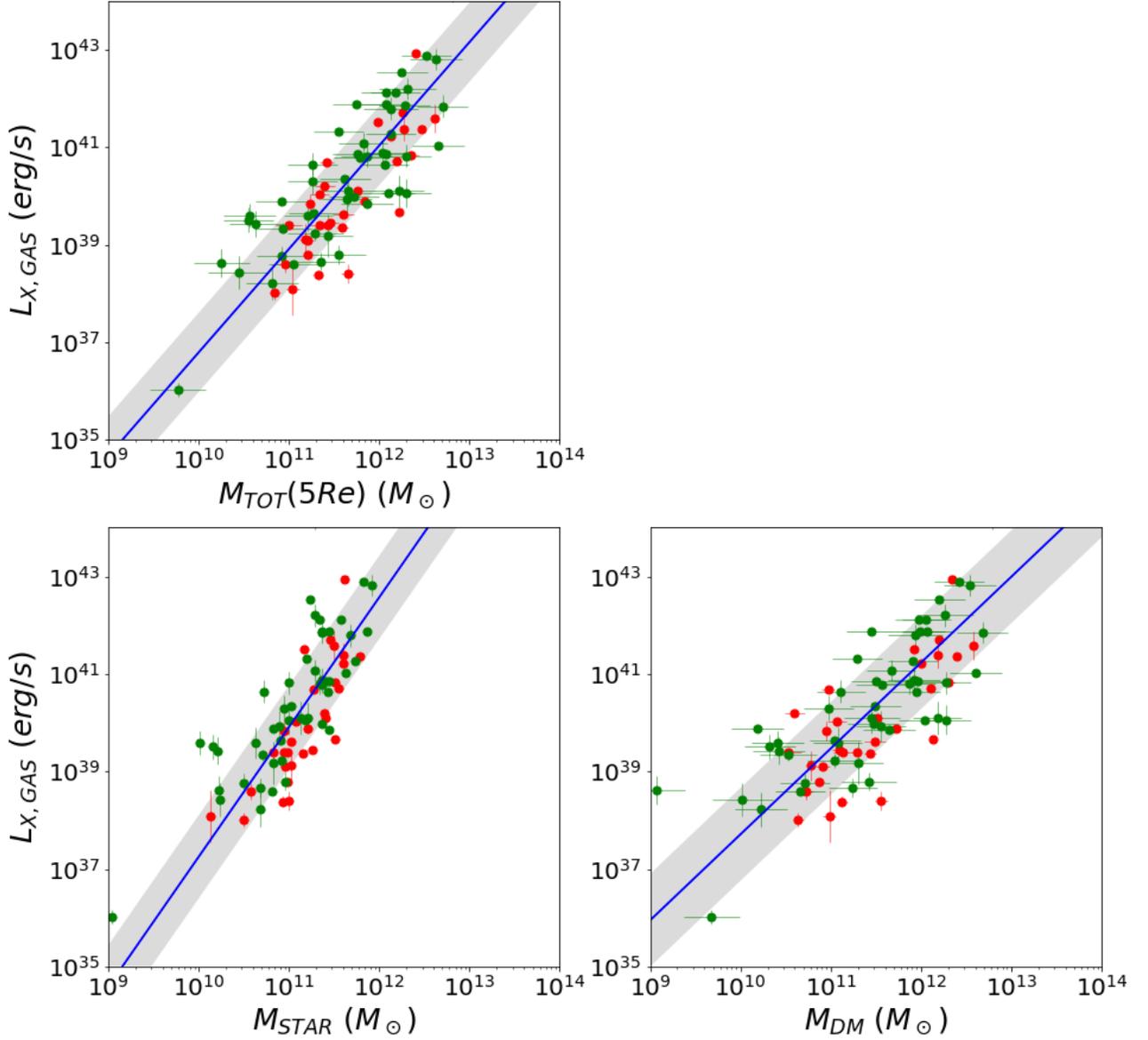

Fig. 9. The $L_{X,GAS}$ − M relationships with (top) $M_{TOT}$ within 5Re, (bottom left) $M_{STAR}$, and (bottom right) $M_{DM}$. The green and red points are for the 30 ETGs with kinematically determined mass measurements (within five effective radii) and the 53 additional galaxies with total masses estimated based on $M_{GCS}$, respectively.

In Figure 8, the most massive galaxies ($M_{TOT}(5Re) \gtrsim 10^{11.5}$ $M_\odot$) have high $f_{DM}$ (i.e., low $f_{STAR}$), while the less massive galaxies exhibit a wider range of $f_{DM}$. Notably, there appears to be a trend among low mass galaxies ($M_{TOT}(5Re) \lesssim 10^{11.5}$ $M_\odot$) with galaxies with high $f_{DM}$ (the red points) being found mostly below the best fit line, while galaxies with higher $f_{STAR}$ or lower $f_{DM}$ (the cyan-blue points) are consistently above the line. These low mass galaxies are essentially the cusp ETGs (Section 3.2), suggesting that stellar mass could be an important secondary factor for determining the amount of the hot ISM among these low $L_{X,GAS}$ galaxies.

To better illustrate the above trend, we explored the effect of using either $M_{TOT}$, $M_{STAR}$ or $M_{DM}$ in the $L_{X,GAS}$ – Mass relation (Figure 9) and measured the deviation in $L_{X,GAS}$ from the best fit lines. These residual $\Delta \log(L_{X,GAS})$ are plotted as a function of $f_{DM}$ in Figures 10 (for the galaxies with kinematically determined $M_{TOT}$) and Figure 11 (for all the galaxies).

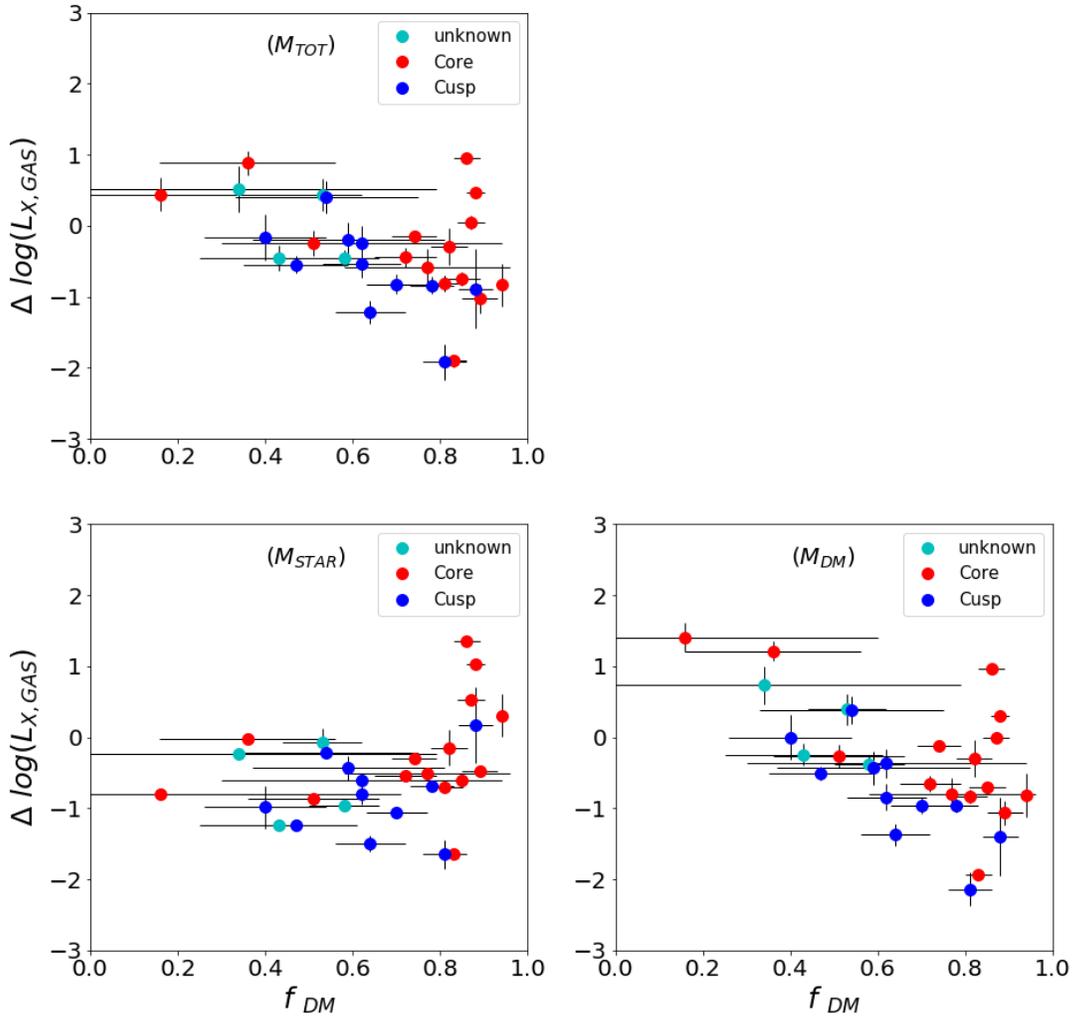

Fig 10. $\Delta \log(L_{X,GAS})$ measured vertically from the best fit $L_{X,GAS}$ – M relation in Figure 9 is plotted against dark matter fraction ($f_{DM} = 1 - f_{STAR}$). $\Delta \log(L_{X,GAS})$ is the relative excess or deficit for a given (top) $M_{TOT}$, (bottom left), $M_{STAR}$, and (bottom right) $M_{DM}$. We show only those galaxies with kinematically determined masses. The core and cusp galaxies are marked by red and blue circles respectively. The unknowns are green circles.

The top panel of Figure 10 shows the excess or deficit in $L_{X,GAS}$, relative to the best fit line of the $L_{X,GAS}$ - $M_{TOT}$ relation, as a function of $f_{DM}$. The cusp and core galaxies are marked by blue and red circles, respectively. The core galaxies do not show a significant trend, except the fact that most of them have high $f_{DM}$. Instead, the cusp subsample appears to have a negative (or positive) relation between $\Delta \log(L_{X,GAS})$ and $f_{DM}$ (or $f_{STAR}$). In other words, for a given $M_{TOT}$, galaxies with higher $M_{STAR}$ have higher $L_{X,GAS}$ (i.e., more hot gas), relative to the best fit relation. The p-value is 0.03 or 0.003 for Pearson and Spearman test, respectively, significantly lower than that (0.1 - 0.2) for the core subsample (see Table 7).

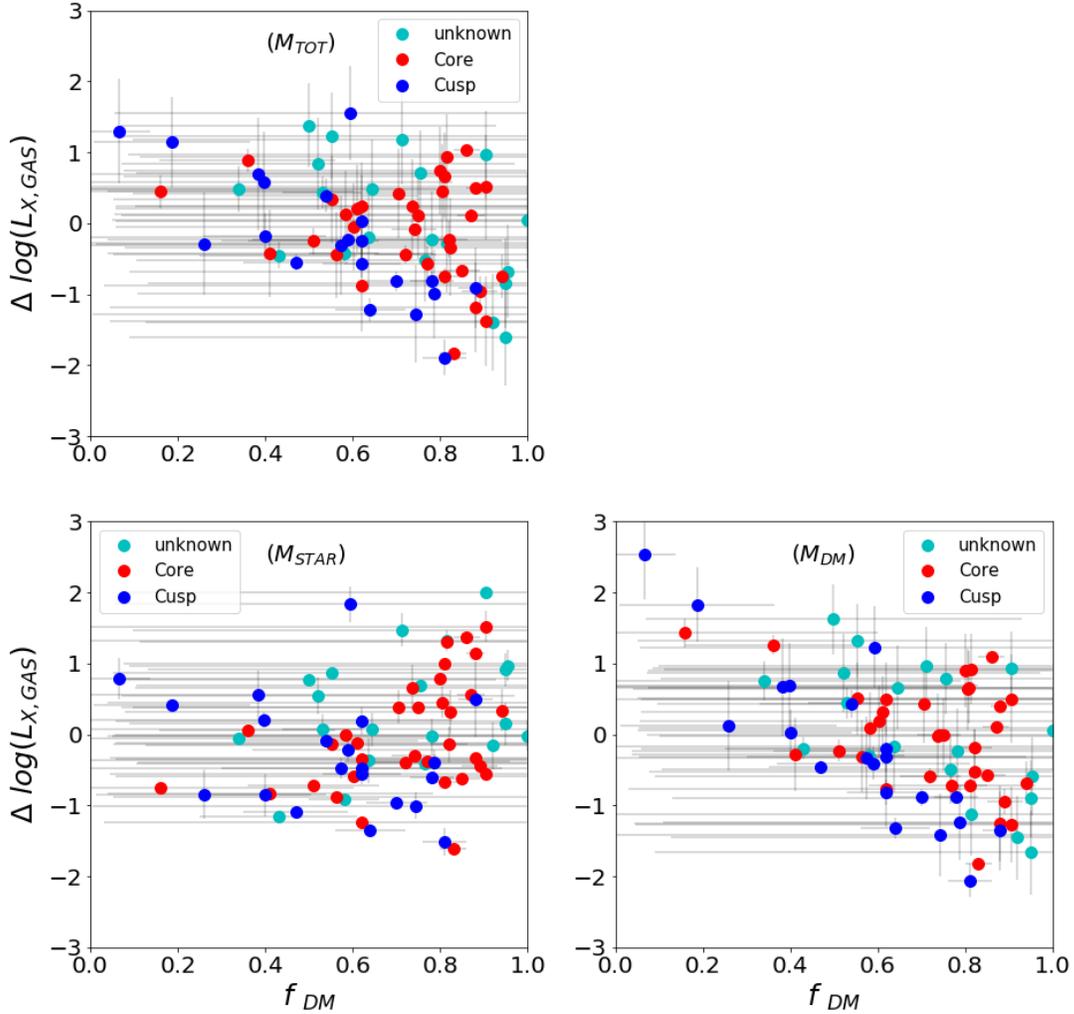

Fig. 11. same as Figure 10, but we also show those galaxies with masses scaled from $M_{GCS}$.

The bottom left and right panels of Figure 10 show the excess of $L_{X,GAS}$, relative to the best fit lines, for a given $M_{STAR}$ and $M_{DM}$, respectively. For the core subsample, again we find less of a definite trend. For the cusp sample, the correlation between $\Delta \log(L_{X,GAS})$ and $f_{DM}$ seen in the top panel disappears in the bottom left panel, but is enhanced in the bottom right panel. These results mean that once $M_{STAR}$ is fixed (the bottom left panel), $L_{X,GAS}$ is not affected by $f_{DM}$, but that once

$M_{DM}$ is fixed (the bottom right panel), $L_{X,GAS}$ is significantly affected by $f_{DM}$, in the sense that galaxies with higher $M_{STAR}$ have higher $L_{X,GAS}$. The p-value is now 0.005 or 0.0005 for Pearson and Spearman test, respectively (see Table 7).

Figure 11 shows the same relations as in Figure 10, but for the entire sample including ETGs with $M_{TOT}$ scaled from $M_{GCS}$. Despite the large errors (in $f_{DM}$), the same behavior is observed. Again, there is no obvious trend, once $M_{STAR}$ is fixed. The most pronounced trend is among cusp galaxies on the bottom left panel. For a given $M_{DM}$, galaxies with higher $M_{STAR}$ (or $f_{STAR}$) have higher $L_{X,GAS}$. The correlation is even stronger in the larger sample with a slope of $5.3 \pm 0.5$ and the corresponding p-value is low, $3\text{-}5 \times 10^{-7}$ (see Table 8). More importantly, this relation now extends to the entire $f_{DM}$ range from 0.1 to 0.9. We note that the leftmost two galaxies (NGC 2434 and NGC 3599) have a small number of GCs and may be subject to a systematic error. Nonetheless, we emphasize that this correlation is seen in the sample with kinematically determined $M_{TOT}$ (in Figure 10) as well as in the full sample (in Figure 11). Even though the $M_{TOT}$-$M_{GCS}$ relation is not well constrained in the cusp subsample, this fact provides an indirect proof for the validity of the $M_{TOT}$ - $M_{GCS}$ relation even in the cusp subsample.

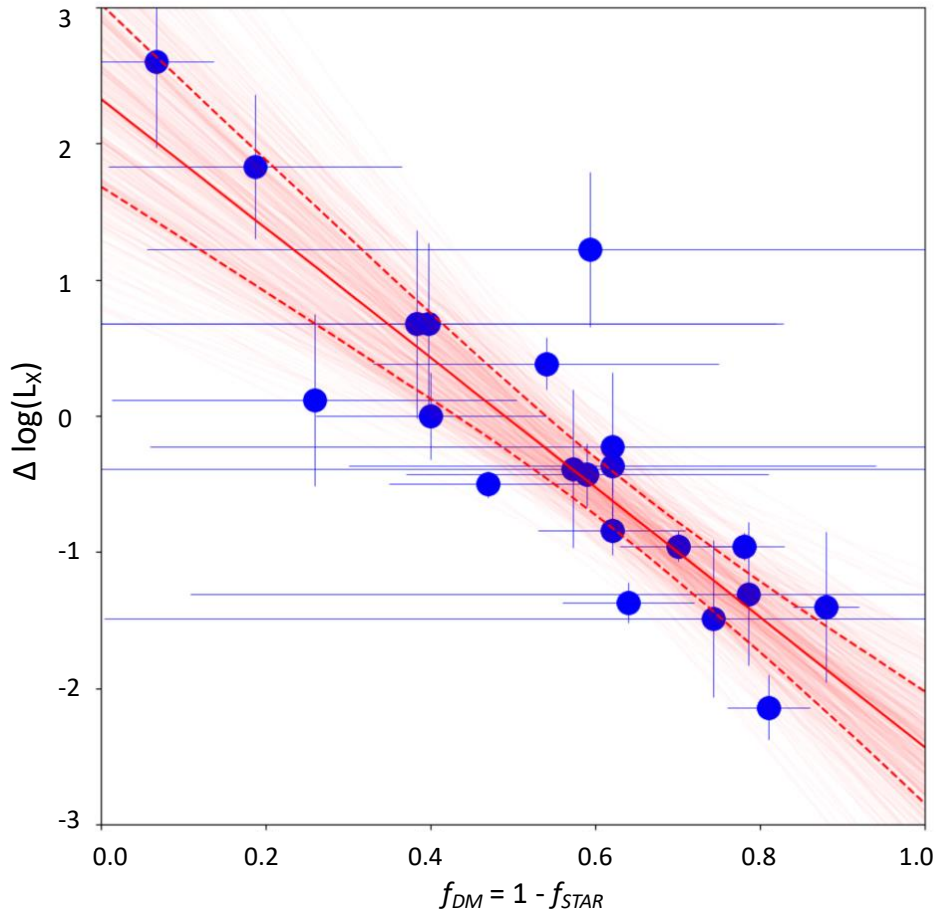

Fig. 12. Residuals of the $L_{X,GAS}$ - $M_{DM}$ correlation as a function of $f_{DM}$ for cusp ETGs. The solid and dashed red lines are the median and $1\sigma$ quantiles from 10,000 Markov chain Monte Carlo simulations.

To further quantify the statistical significance of the above relation in the cusp subsample and to properly consider the large errors of individual data, we apply the Markov chain Monte Carlo (MCMC) method[3] (see Kelly 2007) to assess the allowed range of parameters (see Figure 12). We perform 10,000 MCMC simulations to produce samples from the posterior distribution of the model parameters. The thin red lines in the bottom panel are the Markov chain simulations showing the range of the model parameters. The thick red solid and dashed lines are the median and $1\sigma$ quantiles. Again among the cusp subsample, $L_{X,GAS}$ is higher, relative to the best fit relation, for galaxies with higher $M_{STAR}$ for a given $M_{DM}$.

```
Table 7. Δlog(Lx,GAS) - fDM relation for galaxies with kinematically determined masses
====================================================================================
sample      # of galaxies   slope  error  intercept error    RMS         p-value
                                                            (dex)   pearson  spearman
====================================================================================
for a given MTOT
All              30         -2.79   0.66    1.48   0.41     0.66    0.0121   0.0153
Core             15         -2.72   1.39    1.71   1.07     0.71    0.0993   0.2483
Cusp             11         -3.93   1.36    1.89   0.88     0.47    0.0285   0.0022

for a given MSTAR
All              30          2.33   0.96   -2.04   0.62     0.70    0.0859   0.0466
Core             15          2.27   1.86   -1.89   1.48     0.75    0.2570   0.0498
Cusp             11          1.68   2.53   -1.90   1.57     0.54    0.6397   0.9576

for a given MDM
All              30         -3.70   0.66    2.07   0.41     0.72    0.0009   0.0049
Core             15         -3.75   1.66    2.52   1.31     0.73    0.0127   0.1515
Cusp             11         -4.73   1.27    2.25   0.82     0.45    0.0051   0.0005
====================================================================================

Table 8. Δlog(Lx,GAS) - fDM relation for all galaxies with masses
====================================================================================
sample      # of galaxies   slope  error  intercept error    RMS         p-value
                                                            (dex)   pearson  spearman
====================================================================================

for a given MTOT
All              79         -2.97   0.46    1.95   0.31     0.79    0.0003   0.0003
Core             34         -2.14   1.14    1.45   0.79     0.70    0.1695   0.1968
Cusp             21         -4.04   0.62    2.03   0.43     0.64    0.0002   1.79e-05

for a given MSTAR
All              79          2.05   0.75   -1.34   0.51     0.86    0.0754   0.03937
Core             34          3.15   0.83   -2.29   0.59     0.73    0.0181   0.03037
Cusp             21         -2.45   1.26    1.08   0.74     0.79    0.1631   0.09564

for a given MDM
All              79         -3.89   0.45    2.59   0.30     0.82    4.81e-07 1.20e-05
Core             34         -3.15   0.98    2.24   0.69     0.73    0.01791  0.05841
Cusp             21         -5.30   0.54    2.74   0.37     0.58    5.05e-07 2.50e-07
====================================================================================
```

---

[3] Python version from https://github.com/jmeyers314/linmix

The cusp ETGs have overall low $L_{X,GAS}$, and the observed gas mass is smaller than that produced by stellar outgassing over the entire galaxy lifetime. For typical (hot gas poor, optically small) cusp ETGs with $L_K = 5 \times 10^{10}$ $L_\odot$ and $L_{X,GAS} = 10^{39}$ erg s$^{-1}$ (see Figure 1), the accumulated mass loss in the last 10 Gyr can reach $10^{10}$ $M_\odot$, without counting the first ~2 Gyr when the mass loss could have been even higher (e.g., see Pellegrini 2012). This gas mass is already 1-2 orders of magnitude larger than the observed value. For example, a gas-poor, but optically bright galaxy NGC 1316 ($L_K = 6 \times 10^{11}$ $L_\odot$) has only $M_{GAS} \sim 10^9$ $M_\odot$ (e.g., Kim et al. 1998) and a hot gas rich core galaxy, NGC 4636 ($L_{X,GAS} = 3 \times 10^{41}$ erg s$^{-1}$) has a comparable gas mass, $M_{GAS} \sim 10^{10}$ $M_\odot$ (Trinchieri et al. 1994). Based on these simple comparisons, the hot gas amount in those typical cusp galaxies is comparable to that accumulated for the last ~1 Gyr. Their hot gas is therefore in outflow or even in a wind state (e.g., Ciotti et al. 1991, Pellegrini 2012). In this case, the current rate of gas input from the stellar mass loss, which is proportional to the stellar mass, may become important, suggesting that the current stellar feedback could be an important secondary factor.

## 4.2. Stellar Mass Fraction versus Halo Mass in ETGs

The present understanding of galaxy formation is that each galaxy forms within a dark matter halo, and galaxy formation efficiency is a function of the halo mass, with a peak at halo mass of ~$10^{12}$ $M_\odot$ (see review by Wechsler and Tinker 2018, and their Figure 2). The galaxy formation efficiency (or stellar fraction, $f_{STAR} = M_{STAR} / M_{TOT}$) declines toward both higher and lower mass ends, leading to the conclusion that the efficiency could be considerably suppressed due to AGN feedback at the higher mass galaxies and stellar feedback at the lower mass range (e.g., see the review by Silk & Mamon 2012).

To obtain the total mass ($M_{TOT}$, or halo mass which is dominated by dark matter), and relate it to the stellar mass fraction $f_{STAR}$, various techniques have been used to match observed galaxies and simulated halos. They include abundance matching, halo occupation distribution and some variations of these two methods. While observationally determined galaxy properties (e.g., the galaxy luminosity function) are used to constrain the galaxy-halo matches, they depend on cosmological models and simulations (see Wechsler and Tinker 2018).

The $M_{GCS}$ - $M_{TOT}$ scaling relations (Section 3.1) instead provide a model-independent way to explore the $f_{STAR} - M_{TOT}$ relation. We estimated $M_{TOT}$ for the 242 early-type galaxies with $M_{GCS}$ compiled by Harris et al. (2013) but do not overlap with the Alabi et al. (2017) sample, and we also used the kinematically measured $M_{TOT}$ of Alabi et al. when available. This gives us the largest set of $M_{TOT}$ ever assembled, covering 272 ETGs.

Figure 13a shows this ETG sample plotted over the $f_{STAR} - M_{TOT}$ plot of Behroozi et al. (2013), who used an abundance matching technique to constrain a parametric stellar mass-halo mass relationship. Behroozi et al. (2013) also showed that their results are consistent with those obtained by using the halo occupation distribution. Harris et al. (2013) showed a similar plot (in their Fig. 14) with specific mass $S_M$ (= $M_{GCS} / M_{dyn}$). In Figure 13b we bin our data to improve statistics, by using mass bins of 0.5 dex, and calculating the average stellar mass fraction for each bin. We also compare our estimates with other estimators as indicated in the legend. We use $M_{TOT}(R_{200})$ in the figure, to approximate the virial mass.

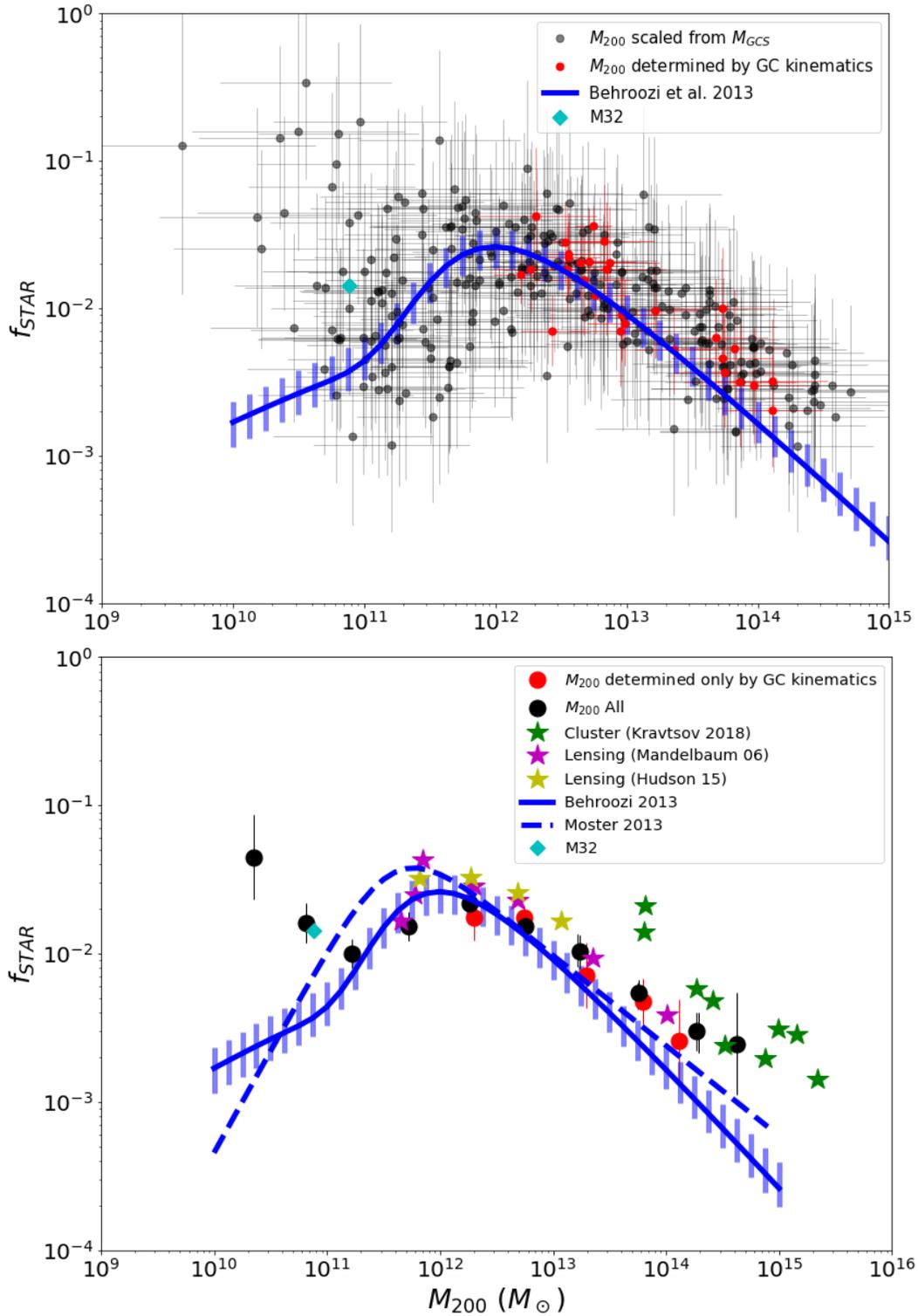

Fig. 13. Stellar and halo mass relation. $f_{STAR}$, the stellar fraction, is plotted against $M_{200}$, the halo mass. In the top panel, the data for individual galaxies are compared with the result from Behroozi et al. (2013). In the bottom panel, the average $f_{STAR}$ (in $M_{200}$ bin of 0.5 dex) is compared with the galaxy cluster and lensing data from the literature.

The peak of our data agrees with Behroozi et al. (2013), giving us added confidence on our use of $M_{GCS}$ to estimate $M_{TOT}$, but the agreement is poorer at higher and lower masses. The average stellar mass fractions in the two lowest mass bins ($M_{200} < 10^{11}$ $M_\odot$) are higher with our estimated $M_{TOT}$, and the discrepancy is significant at the 3-4σ level. We note that our the $M_{TOT}$ - $M_{GCS}$ relation is not verified at the lowest mass bin in section 3. In the 2$^{nd}$ lowest mass bin, M32, which follows the $L_{X,GAS}$ − $M_{TOT}$ relation in Fig. 4, has $f_{STAR}$ almost identical to the mean value. However, it is not clear whether M32, being a relatively rare compact elliptical (cE), represents the low mass system, because M32 would have been affected by the strong tidal force of M31 and most of its outer layers might have been stripped away (e.g., Bekki et al. 2001) and because both measurements of $M_{GCS}$ and $L_{X,GAS}$ are subject to large errors. Interestingly, van Dokkum et al (2018) reported little or no dark matter in the ultra diffuse galaxy, NGC1052-DF2, (with stellar mass ~ 2 x 10$^8$ $M_\odot$), using the kinematics data of 10 GCs associated with the galaxy, claiming that dark matter is not always coupled with baryonic matter on galactic scales. Alternatively, they could be dark matter-free tidal dwarf galaxies (TDGs, e.g., Haslbauer et al. 2019). However, there is also an opposing view on this subject (Martin et al. 2018).

The disagreement at higher masses can already be observed in the Alabi et al. kinematics data alone and is reinforced by the full data set. The discrepancy is marginally significant at the 2-2.5σ level in $M_{200} > 10^{14}$ $M_\odot$. Our results are close to the measurements from the lensing data where they overlap (e.g., Mandelbaum et al. 2006) and the cluster mass estimations by Kravtsov et al. (2018) show the same trend to even higher masses.

In summary, our results, which are independent of cosmological models, show an overall agreement with those determined by abundance matching and halo occupation methods. However, we obtain higher stellar mass fractions at the lower and higher masses, suggesting that feedback may not be as strong as generally believed in these regimes.

## 5. SUMMARY AND CONCLUSIONS

1) We have assembled a sample of 83 ETGs, to explore the use of the mass of their GC systems $M_{GCS}$ as a proxy for the total galaxy mass $M_{TOT}$. We first examined the scaling relation between $M_{TOT}$ and $M_{GCS}$ by using kinematically determined $M_{TOT}$ (within 5 Re) in a sample of 30 ETGs, finding an excellent quasi-linear correlation. We then extended the comparison to other recently established proxies of $M_{TOT}$: the X-ray luminosity of the hot ISM in these galaxies, and the temperature of the ISM (Kim & Fabbiano 2013; Forbes et al. 2017). Examining the $L_{X,GAS}$ - $M_{TOT}$ relation in a sample of 83 ETGs and the $T_{GAS}$ - $M_{TOT}$ relation in a sample of 57 ETGs, we confirm that $M_{GCS}$ can be effectively used as a proxy of $M_{TOT}$. The relation between $M_{GCS}$ and $M_{TOT}$ is near-linear:

   $\log (M_{TOT}(5Re) / 10^{11.7}$ $M_\odot) = 0.85\pm0.06$ x $\log (M_{GCS} / 10^{8.5}$ $M_\odot)$
   or
   $\log (M_{TOT}(R_{200}) / 10^{13.1}$ $M_\odot) = 0.99\pm0.07$ x $\log (M_{GCS} / 10^{8.5}$ $M_\odot)$

2) We find that the $L_{X,GAS}$ − $M_{TOT}$ and $T_{GAS}$ − $M_{TOT}$ relations are tighter in the core ETGs, compared with the cusp ETGs in our sample. These differences suggest that $M_{TOT}$ is the most important factor in retaining hot ISM in the core subsample (mostly old, passively evolving stellar systems), while a secondary factor may be at play in the cusp subsample. These conclusions are consistent with those of our previous work, based on smaller galaxy samples (Kim & Fabbiano 2015, Forbes et al. 2017).

3) For each ETG we estimate the integrated stellar mass $M_{STAR}$ from the K-band luminosity assuming $M_{STAR}/L_K = 1$ $M_\odot/L_\odot$. Since in our ETG sample the gas mass is one or two orders of magnitude smaller than $M_{STAR}$, we can write $M_{TOT} = M_{DM} + M_{STAR}$ and $f_{DM} + f_{STAR} = 1$, from which we estimated $f_{DM} = M_{DM}/M_{TOT}$. We find that the more massive ETGs ($M_{TOT}(5Re)$ $\gtrsim 10^{11.5}$ $M_\odot$) have high $f_{DM}$ (i.e., low $f_{STAR}$), while the less massive ones (mostly cusp ETGs) exhibit a wider range of $f_{DM}$. We further find that the cusp ETGs with higher stellar mass (or $f_{STAR}$) for a given $M_{DM}$ (or $M_{TOT}$) have higher values of $L_{X,GAS}$. However, once $M_{STAR}$ is fixed, no obvious relation exists as a function of $M_{DM}$ (or $f_{DM}$). These trends suggest that on-going (or recent) stellar feedback could be an important secondary factor for determining the amount of the hot ISM in low $L_{X,GAS}$ cusp galaxies.

4) With 272 ETGs with $M_{TOT}$ (scaled from $M_{GCS}$ by the above relation), we investigate the relation between $M_{TOT}$ and $f_{STAR}$, we find overall agreement with previous results using other methodologies (e.g., abundance matching), i.e., the stellar mass fraction peaking at $M_{TOT} \sim 10^{12}$ $M_\odot$ and declining toward both higher and lower mass ends. However, we identify that the stellar fraction is quantitatively larger than the previously known relation both at lower and higher mass ends, indicating the star formation efficiency is less suppressed than previously expected at the higher and lower mass range.

## ACKNOWLEDGEMENT

We thank Gretchen Harris and Bill Harris for their encouraging discussions during the initial phase of this work and sending their GC data. We have used the CXC CIAO software package, the NASA NED and ADS facilities. This work was supported by the Chandra GO grant AR5 16007X, 2018 Smithsonian Scholarly Study and NASA contract NAS8-03060 (CXC). DAF thanks the ARC for financial support via DP160101608.

Table 1.

```
--------------------------------------------------------------------------------
 name    T      d    logL_K  core   ref(X)  logL_X,GAS  T_GAS    log M_GCS  ref(GC)    logM_TOT(5Re)
                                                                                    from GC_kin   from M_GCS
 (1)    (2)    (3)    (4)    (5)     (6)       (7)       (8)       (9)      (10)      (11)          (12)
--------------------------------------------------------------------------------
N0221   -6     0.76   9.04          BKF11   36.02 0.15             6.26 .10   53                   9.77 .31
N0474   -2    29.51  10.83  Int     KF15    39.18 0.43             8.22 .06   39,48               11.43 .28
N0524   -1    23.13  11.18  Core    KF15    40.06 0.06  0.50 .07   9.01 .11   32,39               12.10 .29
N0541   -3    73.10  11.34          OFP01   40.80 0.22             8.73 .11   35                  11.87 .29
N0708   -5    47.07  11.23          G16     42.53 0.01  1.91 .07   9.18 .08   6,41                12.25 .28
N0720   -5    27.67  11.29  Core    BKF11   40.71 0.01  0.54 .01   8.32 .11   37      11.41 .08
N0821   -5    23.38  10.89  Cusp    KF15    38.40 0.19             7.97 .06   49      11.65 .08
N1023   -3    11.43  10.93  Cusp    KF15    38.81 0.03  0.30 .02   8.14 .03   55      11.20 .05
N1052   -5    19.35  10.90  Core    BKF11   39.64 0.03  0.34 .02   8.04 .05   22                  11.28 .28
N1132  -4.5   96.09  11.57          G16     42.12 0.07  1.09 .10   9.10 .03   1                   12.18 .27
N1316   -2    21.09  11.73  Core    BKF11   40.72 0.01  0.60 .01   8.57 .18   26,54   12.20 .05
N1332   -3    22.91  11.21          OPC03   40.10 0.32  0.41 .05   8.47 .18   39                  11.65 .31
N1374  -4.5   19.64  10.63  Core    OFP01   39.59 0.32             7.96 .02   3                   11.21 .27
N1387   -3    19.82  10.94          OFP01   40.30 0.26             8.02 .03   3                   11.26 .27
N1399   -5    20.68  11.42  Core    OPC03   41.44 0.34  1.21 .03   9.28 .05   2,49    12.61 .03
N1400   -3    26.42  11.03  Cusp    OFP01   39.98 0.33             8.93 .12   43      11.34 .12
N1404   -5    20.43  11.20          OPC03   41.31 0.05  0.60 .01   8.36 .07   21,27,46            11.55 .28
N1407   -5    28.84  11.55  Core    OPC03   41.36 0.25  0.79 .08   9.40 .11   50      12.27 .04
N1427  -4.1   20.52  10.68  Cusp    BKF11   38.66 0.19  0.38 .18   8.12 .05   23,38               11.35 .27
N1549   -5    19.68  11.19          BKF11   39.49 0.06  0.35 .04   7.63 .15   9                   10.93 .30
N1600   -5    66.00  11.73  Core    G16     41.26 0.02  1.10 .05   9.04 .07   47                  12.13 .28
N1700   -5    44.26  11.37  Core    G16     40.78 0.02  0.34 .04   8.63 .08   11                  11.78 .28
N2434   -5    21.58  10.83  Cusp    BKF11   39.88 0.03  0.52 .02   7.61 .11   25                  10.92 .29
N2768   -5    22.39  11.21  Cusp    KF15    39.90 0.02  0.31 .01   8.36 .12   39      11.83 .06
N2778   -5    22.91  10.23  Cusp    KF15    38.42 0.34  0.54 .40   7.05 .20   25                  10.44 .33
N2832   -4   102.80  11.86  Core    G16     41.86 0.08  1.13 .12   9.22 .09   6,35                12.28 .28
N3115   -3    10.00  10.96  Cusp    BKF11   38.44 0.11  0.44 .13   8.19 .06   19      11.32 .06
N3258   -5    32.06  11.00          OFP01   40.82 0.23             9.24 .01   4                   12.30 .27
N3268   -5    34.83  11.13          OFP01   40.10 0.33             9.15 .01   4                   12.22 .27
N3311   -4    44.26  11.36          OFP01   41.84 0.22             9.71 .05   17                  12.70 .28
N3377   -5    11.04  10.42  Cusp    KF15    38.01 0.13  0.19 .13   7.66 .03   15      10.85 .06
N3379   -5    10.20  10.82  Core    KF15    38.60 0.07  0.25 .02   7.77 .08   19,45               11.05 .28
N3384   -3    10.80  10.68  Cusp    KF15    38.22 0.35             7.49 .10   29                  10.81 .29
N3414   -2    25.23  10.92  Cusp    KF15    39.22 0.14  0.57 .18   8.05 .18   39                  11.29 .31
N3585   -5    21.20  11.28  Int     BKF11   39.22 0.08  0.36 .05   7.96 .12   44                  11.21 .29
N3599   -2    20.32  10.22  Cusp    KF15    38.62 0.29  0.16 .13   6.82 .13   39                  10.25 .31
N3607   -2    20.00  11.12  Core    KF15    40.11 0.05  0.59 .07   8.25 .12   38      11.40 .12
N3608   -5    23.00  10.80  Core    KF15    39.64 0.07  0.40 .07   8.10 .16   38      11.60 .13
N3842   -5    94.90  11.63  Core    G16     41.03 0.06  1.32 .11   9.65 .10   6,12                12.65 .29
N3923   -5    22.91  11.43          BKF11   40.64 0.01  0.45 .01   8.96 .04   19,48,56            12.06 .27
N4073  -3.8   89.00  11.82  Core    G16     42.88 0.07  1.85 .04   9.50 .03   10,                 12.52 .28
N4203   -3    15.14  10.71  Cusp    KF15    39.34 0.11  0.25 .08   7.63 .20   39                  10.93 .32
N4261   -5    31.62  11.41  Core    KF15    40.86 0.01  0.76 .01   8.60 .08   8                   11.76 .28
N4278   -5    16.07  10.85  Core    KF15    39.41 0.02  0.30 .01   8.48 .10   33,38   11.43 .06
N4340   -1    16.00  10.39  Cusp    OFP01   39.38 0.31             6.91 .13   42                  10.32 .31
N4365   -5    23.33  11.39  Core    KF15    39.67 0.02  0.46 .02   9.03 .07   42      12.22 .04
N4374   -5    18.51  11.36  Core    KF15    40.82 0.08  0.73 .01   9.15 .11   42      12.35 .09
N4382   -1    17.88  11.36  Core    KF15    39.99 0.02  0.39 .02   8.56 .07   42                  11.72 .28
N4387   -5    18.00  10.16  Cusp    OFP01   39.51 0.25             7.18 .06   42                  10.55 .28
N4406   -5    17.09  11.34  Core    KF15    42.12 0.00  0.82 .01   8.97 .03   42                  12.07 .27
N4458   -5    16.32  10.01  Int     OFP01   39.59 0.24             7.19 .07   42                  10.56 .29
N4459   -1    16.01  10.86  Cusp    KF15    39.39 0.08  0.40 .11   7.76 .05   42      11.34 .12
N4472   -5    17.03  11.62  Core    KF15    41.38 0.04  0.95 .01   9.39 .05   42      12.47 .04
N4473   -5    15.25  10.82  Core    KF15    39.10 0.07  0.31 .03   8.00 .10   42      11.20 .08
N4486   -4    17.00  11.45  Core    KF15    42.95 0.00  1.50 .00   9.65 .03   42      12.41 .03
N4494   -5    17.06  10.98  Cusp    KF15    39.14 0.27  0.34 .30   8.05 .05   24      11.18 .06
N4526   -2    16.90  11.18          KF15    39.47 0.03  0.31 .02   8.06 .11   42      11.46 .09
N4550  -1.5   15.44  10.21  Int     OFP01   39.43 0.27             7.27 .09   42                  10.63 .29
N4552   -5    15.89  11.02  Core    KF15    40.34 0.01  0.59 .01   8.44 .07   42                  11.62 .28
N4564   -5    15.87  10.54  Cusp    KF15    38.58 0.17             7.71 .06   42      10.95 .10
N4589   -5    21.98  10.89  Core    OFP01   39.93 0.31             8.46 .06   38                  11.64 .28
N4594    1    9.77   11.31          Other   39.32 0.03  0.60 .01   8.78 .04   45,49   11.76 .05
N4621   -5    14.85  10.96  Cusp    KF15    38.80 0.19  0.27 .07   8.36 .16   42                  11.55 .30
N4636   -5    14.66  11.08  Core    KF15    41.52 0.01  0.73 .01   9.09 .01   18      11.98 .03
N4649   -5    17.09  11.48  Core    KF15    41.25 0.04  0.86 .00   9.13 .05   42      12.13 .03
N4697   -5    12.01  10.92  Cusp    KF15    39.32 0.02  0.31 .01   7.81 .12   36      11.59 .07
N4762   -2    23.88  11.15  Cusp    OFP01   40.15 0.32             7.86 .04   29,42               11.13 .28
N4889   -4    96.60  11.92  Core    OFP01   42.81 0.22             9.63 .07   6,34,40             12.63 .29
```

```
N5128 -2     3.80 10.91        Other 40.20 0.03 0.29 .20 8.57 .09 30           11.23 .05
N5193 -4.5 37.99 11.00         OFP01 40.06 0.30           9.24 .03 41                      12.30 .28
N5322 -5    31.19 11.44 Core   KF15  39.85 0.06 0.33 .04 8.72 .12 17                      11.86 .29
N5813 -5    32.21 11.36 Core   KF15  41.87 0.00 0.70 .01 8.98 .06 29                      12.08 .28
N5845 -4.6 25.94 10.50 Cusp    KF15  38.77 0.20 0.39 .21 7.62 .07 38                      10.92 .28
N5846 -5    24.89 11.32 Core   KF15  41.73 0.01 0.72 .01 9.17 .10 20           12.26 .04
N5866 -1    15.35 10.94        KF15  39.43 0.04 0.32 .02 8.01 .08 13,29        11.00 .17
N5982 -5    41.47 11.29 Core   OFP01 41.07 0.23           8.67 .05 48                      11.82 .27
N6173 -5   125.80 11.29 Core   OFP01 42.20 0.22           9.25 .19 6                       12.31 .32
N6482 -5    55.14 11.45        G16   41.88 0.07 0.71 .04 8.59 .03 1                        11.75 .27
N7049 -2    29.92 11.37        OFP01 40.87 0.25           8.92 .11 17                      12.03 .29
N7173 -4.1 31.33 10.72         OFP01 40.64 0.23           8.01 .03 15                      11.26 .27
N7457 -3    13.24 10.28 Cusp   KF15  38.10 0.58           7.56 .12 14,16,28 11.04 .08
N7626 -5    47.42 11.45 Int    G16   40.85 0.03 0.71 .05 8.98 .04 48                      12.08 .27
N7768 -5   112.10 11.68        OFP01 41.79 0.23           9.04 .16 6,35                    12.13 .30
-----------------------------------------------------------------------------------------------------
```

(1) galaxy name
(2) morphological type (T) from RC3
(3) distance in Mpc
(4) log $L_K$ (in $L_\odot$)
(5) nuclear profile (core, cusp, intermediate)
(6) reference of the X-ray data in columns 7 and 8
(7) log $L_{X,GAS}$ (in erg/s) and error
(8) $T_{GAS}$ (in keV) and error
(9) log $M_{GCS}$ and error from Harris (2013, 2017)
(10) reference of the GCS data
(11) log $M_{TOT}$ (r < 5 $R_e$) and error in $M_\odot$ measured from GC kinematics (from Alabi et al. 2017)
(12) log $M_{TOT}$ (r < 5 $R_e$) and error in $M_\odot$ scaled from $M_{GCS}$

### X-ray references

### GC references